\documentstyle[preprint,aps,prbbib]{revtex}                                                 

\begin{document}     

\begin{center}
{\Large\bf Cosmic Temperature Decline in the Course of the Evolution of the
Universe}

\end{center}

\begin{center}
{Silvia Behar and Moshe Carmeli}
\end{center}

\begin{center}
{Department of Physics, Ben Gurion University, Beer Sheva 84105, 
Israel}
\end{center}

\begin{center}
{Email: carmelim@bgumail.bgu.ac.il}
\end{center}

\begin{abstract}
At the early stage of the Universe-evolution there were no stars and no 
galaxies, but only a uniform hot plasma consisting of free electrons and free
nuclei. The Universe temperature was determined by the Stefan-Boltzmann law of
thermodynamics and the general relativistic cosmological theory. At the 
present time one has the background cosmic radiation with the temperature of 
2.73K. We calculate how much of the early Universe energy has gone to matter
and other forms of energy, so as to leave us with a background radiation of
only 2.73K.
\end{abstract}
\newpage
\section{Introduction}
At the early stage of the Universe-evolution there were no stars and no 
galaxies, but only a uniform hot plasma consisting of free electrons and free
nuclei. At very early times, the violent thermal collisions would have 
prevented the existence of any kind of nucleus, and the matter in the Universe
must have been in the form of free electrons, protons and neutrons. The 
temperature of the Universe is related to its evolution and, at the early
stage, is given by a well-known formula which shows that $T\propto t^{-1/2}$.
The question raised here is how to relate this temperature at the
very early stage of the Universe to the present time cosmic background 
temperature 2.73K. Obviously part of the energy that the
Universe had at the early stage has gone to the matter, and the background
radiation temperature represents the other part. In the following it is shown 
that
the ratio of energy that goes to matter to that of the background radiation
is about 13. In other words, if the Universe today would have no matter, the
background cosmic temperature would be about 35K (13$\times$2.73K).
\section{Temperature Formula in the Absence of Gravity}
Our starting point is the familiar thermodynamical formula that relates the
temperature to the cosmic time with respect to the Big Bang [1,2]:
$$T=\left(\frac{45\hbar^3}{32\pi^3k^4G}\right)^{1/4}t^{-1/2}.\eqno(1)$$
In this equation $k$ is Boltzmann's constant and $G$ is Newton's gravitational
constant. Our aim is to transform this temperature to the present time. This
can be done by using the cosmological transformation [3]
$$T=\frac{T_0}{\left(1-\tilde{t}^2/\tau^2\right)^{1/2}},\eqno(2)$$
where $T_0$ is the present time background temperature, $T_0=2.73K$, 
$\tilde{t}$ is the cosmic time measured with respect to us now, and $\tau$ is
the Hubble time in the absence of gravity, $\tau=12.486$Gyr [4,5]. Since we
are looking for temperatures $T$ at very early times, we can use the 
approximation $\tilde{t}\approx\tau$, thus
$$1-\tilde{t}^2/\tau^2=\left(1+\tilde{t}/\tau\right)\left(1-\tilde{t}/\tau
\right)\approx 2\left(1-\tilde{t}/\tau\right)=\left(2/\tau\right)\left(\tau-
\tilde{t}\right)=2t/\tau,\eqno(3)$$
where $t$ is the cosmic time with respect to the Big Bang. Using this result 
in Eq. (2) we obtain 
$$T=T_0\left(\tau/2\right)^{1/2}t^{-1/2},\eqno(4)$$
with the same dependence on time as in Eq. (1).
\section{Comparison}
As is seen, both equations (1) and (4) show that the temperature $T$ depends
on $t^{-1/2}$. The coefficients appearing before the $t^{-1/2}$, however, are
not identical. A simple calculation shows
$$\left(\frac{45\hbar^3}{32\pi^3k^4G}\right)^{1/4}=1.52\times 10^{10}
Ks^{1/2}\eqno(5)$$
for the coefficient appearing in Eq. (1), and
$$T_0\left(\tau/2\right)^{1/2}=1.16\times 10^9 Ks^{1/2},\eqno(6)$$
for that appearing in Eq. (4). In the above equations we have used
$$\hbar=1.05\times 10^{-34}Js,$$
$$k=1.38\times 10^{-23}J/K,$$
$$G=6.67\times 10^{-11} m^3/s^2Kg,\eqno(7)$$
$$T_0=2.73K,$$
$$\tau=12.486Gyr.$$

Accordingly we can write for the temperatures in both cases 
$$T\approx 1.5\times 10^{10}Ks^{1/2}t^{-1/2},\eqno(8)$$
and
$$T\approx 1.2\times 10^9 Ks^{1/2}t^{-1/2}.\eqno(9)$$
The ratio between them is approximately 13.
\section{Conclusion}
It thus appears that the dominant part of the plasma energy of the early
Universe has gone to the creation of matter appearing now in the Universe, and
only a small fraction of it was left for the background cosmic radiation.

\end{document}